\documentclass[reprint,amsmath,amssymb,aps,pre,longbibliography,showkeys]{revtex4-1}

\usepackage{graphicx}
\usepackage{dcolumn}
\usepackage{bm}
\usepackage{color}
\usepackage[utf8]{inputenc}
\usepackage{cancel}

\newcommand{\vX}{\vec{X}}
\newcommand{\vY}{\vec{Y}}

\newcommand{\vJ}{\vec{J}}
\newcommand{\vF}{\vec{F}}

\newcommand{\vXr}{\vec{X}_{\rm R}}

\newcommand{\vXl}{\vec{X}_{\rm L}}
\newcommand{\vXm}{\vec{X}_{\rm M}}

\newcommand{\vd}{{\vec{\delta}}}

\newcommand{\tS}{\tilde{S}}

\newcommand{\tSe}{\tilde{S}^{\rm e}}
\newcommand{\Se}{S^{\rm e}}

\begin{document}

\title{A general thermodynamic approach for diffusion on a lattice}

\author{M. A. Di Muro}
\author{M. Hoyuelos}
\email{hoyuelos@mdp.edu.ar}

\affiliation{Instituto de Investigaciones F\'isicas de Mar del Plata (IFIMAR-CONICET), Departamento de F\'isica, Facultad de Ciencias Exactas y Naturales, Universidad Nacional de Mar del Plata, Funes 3350, 7600 Mar del Plata, Argentina}

\begin{abstract}
This work presents a general thermodynamic approach to describe particle diffusion on a lattice, a model used to study transport processes in solids and on surfaces. By treating each lattice site as an open thermodynamic system, the effects of microscopic particle interactions are represented through the chemical potential. A fundamental relationship between the Onsager matrix ($L$) and its ideal-system counterpart ($L_\text{id}$, where interactions are neglected) using the determinant of the covariance matrix is demonstrated. This framework allows for the calculation of transport coefficients using the combination of their ideal values and thermodynamic properties. The general result is successfully applied to reproduce the Darken equation for substitutional diffusion in solids and to derive the non-diagonal diffusion matrix of the Zhdanov model for surface diffusion of Langmuir particles. In the last case, analytical predictions are further validated through numerical simulations across various interaction potentials.
\end{abstract}


\maketitle

\section{Introduction}

Particle diffusion on a perfect lattice is a theoretical model widely used as a first step to describe diffusion processes in real solids \cite{paul,mehrer,shewmon}, or on surfaces \cite{ala,gomer,antczak,tsong}. 
The main ingredients of the traditional approach are the
assumption of discrete adsorption sites (characterized by the lattice spacing of the crystal), potential barriers, and the frequency of molecular vibration.

From a thermodynamic perspective, microscopic details are ignored and interaction effects are represented by the chemical potential. This approach also utilizes a lattice, which should not be confused with the underlying microscopic crystalline structure. In this framework, each lattice site (or ``cell'') contains many particles and is analyzed as a thermodynamic system; the surrounding sites act as a reservoir that fixes intensive parameters, such as temperature $T$ and chemical potential $\mu$. Spatial variations of thermodynamic quantities are assumed to be smooth compared with the lattice size. 

This thermodynamic approach has successfully described diffusion in solid binary alloys composed of species $A$ and $B$. Darken \cite{darken} established a relationship between the intrinsic diffusion coefficient $D_A$ and the tracer diffusion coefficient $D_A^*$ for species $A$ through the thermodynamic factor $\Gamma$, such that $D_A^* = D_A/\Gamma$. We emphasize the fundamental role of the thermodynamic factor, which is directly related to particle number fluctuations, in determining the density dependence of the diffusion coefficients. This thermodynamic, or coarse-grained, picture was also employed in the kinetic approach presented in Ref.\ \cite{krylov}, which explains complex behaviors such as anomalous prefactors and non-Arrhenius temperature dependence. Another example is the model for surface diffusion of adsorbed species proposed in \cite{zhdanov0,zhdanov}; while a microscopic lattice-gas description defines the rules of movement, the model relies on a thermodynamic description to define the macroscopic transport of species (see also \cite{tammaro,gorban,skakauskas}).

Transport coefficients are inherently linked to thermodynamic fluctuations, which in turn depend on the choice of ensemble. The ensemble considered here is that of an open system at constant volume, characterized by the exchange of conserved quantities with a reservoir. Following the framework introduced in \cite{dimuro}, we demonstrate a simple relationship between the Onsager matrix, $L$, and its ideal counterpart, $L_\text{id}$ (where particle interactions are neglected), through the determinant of the covariance matrix. 
Interactions are thermodynamically encoded in the chemical potential as a function of density and temperature; the only assumption about their nature is that their range is much smaller than the cell size. While Green-Kubo relations \cite{green,kubo} exist for such transport coefficients, they typically require evaluation via molecular dynamics simulations, which often precludes obtaining explicit expressions in terms of thermodynamic properties.

We apply our results to scenarios characteristic of diffusion in solids and on surfaces. In one case, we reproduce the aforementioned Darken equation; in the other, we obtain the diffusion matrix of the Zhdanov model for surface diffusion of two species of Langmuir particles \cite{zhdanov0,zhdanov}.

The paper is organized as follows. In Sec.\ \ref{s.basic} we establish the system definition, notation and fundamental equations. The subsequent three sections develop a general theoretical framework for determining transport coefficients in an ensemble where extensive quantities (excluding volume) fluctuate according to Einstein's fluctuation theory. Specifically, Sec.\ \ref{s.pratio} employs detailed balance to determine the probability ratio of initial and final states during the transport of a conserved quantity between neighboring cells; transition rates are derived in Sec.\ \ref{s.tr}; and Sec.\ \ref{s.transp} utilizes these results to connect transport coefficients to those of an ideal system. These analytical results are then applied to the diffusion of two species in solids and on surfaces in Sec.\ \ref{s.diffusion}. Numerical results confirming our theoretical predictions are presented in Sec.\ \ref{s.numerical}, followed by our conclusions in Sec.\ \ref{s.conclusions}.

\section{System definition and basic equations}
\label{s.basic}

Let us consider a system composed by two cells, each with size $a$ and volume $V=a^3$, identified by sub-indices R and L, in contact with a reservoir. The state of the system is defined by a vector of $n$ extensive quantities for each cell, $\vX_\alpha = (X_{\alpha 1}, \cdots X_{\alpha n})$, where $\alpha=$ R or L; volume $V$ is constant and is not included in the vector $\vX_\alpha$. The macrostate of a single cell, $\vX_\alpha$, fluctuates around the thermodynamic average $\vX$ (without sub-index), which is the same for both cells since the influence of the reservoir is uniform. Entropic intensive parameters $Y_i = \frac{\partial S(\vX)}{\partial X_i}$, where $S(\vX)$ is the thermodynamic entropy, are determined by the reservoir. For example, the components of $\vX_\alpha$ may represent particle number and internal energy, such that $\vX_\alpha = (U_\alpha,N_\alpha)$ and $\vY = (\frac{1}{T},-\frac{\mu}{T})$, where $\mu$ and $T$ denote the chemical potential and temperature. The components of $\vX_\alpha$ represent quantities that can be exchanged between the two cells.

\begin{figure}
	\includegraphics[width=\linewidth]{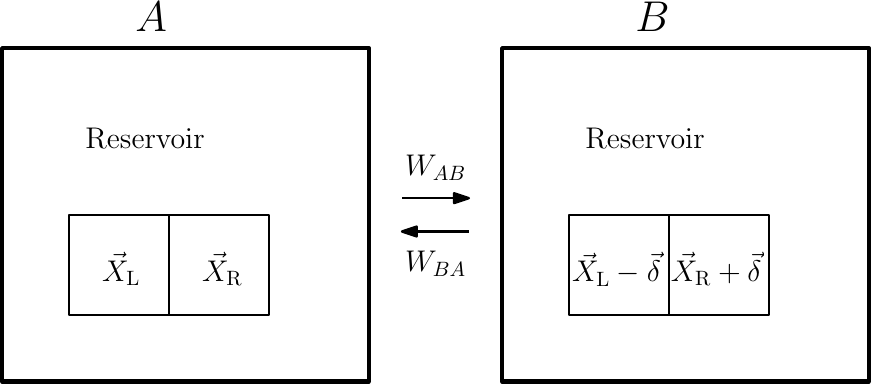}
	\caption{Scheme of the transitions $A\rightarrow B$ and $B\rightarrow A$ where a quantity $\vd$ is transported between cells L and R.}
	\label{f.scheme}
\end{figure}

We analyze transport processes within the framework of classical irreversible thermodynamics, assuming that the characteristic lengths of spatial variations in the state variables are much larger than $a$ and that local thermal equilibrium holds. This implies that transition rates associated to exchange processes between cells R and L must satisfy detailed balance. Consider a configuration where cells L and R are initially in the macrostates $\vXl$ and $\vXr$. A small quantity $\vd$ is transported from L to R, resulting in the final states $\vXl - \vd$ and $\vXr + \vd$. Let $A = (\vXl, \vXr)$ and $B = (\vXl - \vd, \vXr + \vd)$ denote the initial and final states, respectively, as illustrated in Fig.\ \ref{f.scheme}.
The transition rates from $A$ to $B$ and from $B$ to $A$, denoted by $W_{AB}$ and $W_{BA}$, satisfy the detailed balance condition:
\begin{equation}\label{detbal}
	P_A W_{AB} = P_B W_{BA},
\end{equation}
where $P_A$ and $P_B$ are the probabilities of states $A$ and $B$. When interactions at the contact wall between particles in different cells are negligible compared to bulk interactions, these probabilities can be factorized as $P_A = P_{\vXl} P_{\vXr}$ and $P_B = P_{\vXl - \vd} P_{\vXr + \vd}$. This constitute a reasonable approximation for systems containing a large number of particles and with an interaction range much smaller than $a$.

The probability of finding cell $\alpha$ in state $\vX_\alpha$ is given by:
\begin{equation}\label{PX}
	P_{\vX_\alpha} = \exp\left(-\tilde{S}[\vY] + \tilde{S}(\vX_\alpha) - \vY\cdot \vX_\alpha\right),
\end{equation}
where $\tilde{S}[\vY]$ is the statistical mechanics (SM) Massieu function given by $\tilde{S}[\vY] = \ln \mathcal{Z}(\vY)$, and $\mathcal{Z}$ is the partition function for the ensemble with $\vY$ as its natural variables. (Variables in square brackets indicate that a mathematical transformation has been performed on the entropy). The SM entropy is $\tilde{S}(\vX_\alpha) = \ln \Omega(\vX_\alpha)$, where $\Omega(\vX_\alpha)$ is the number of microscopic states consistent with $\vX_\alpha$; see, for example, \cite{callen}. To simplify the notation, we set $k_B=1$, such that temperature is expressed in energy units. Quantities denoted with a tilde are defined within the context of statistical mechanics and contain non-extensive terms. When non-extensive terms are neglected, these quantities are written without tilde and correspond to thermodynamic (TH) functions. This distinction is necessary to correctly account for the non-extensive terms that are relevant when calculating transition probabilities. While TH Massieu functions are obtained from a sequence of Legendre transforms starting from the TH entropy, SM Massieu functions are generated by Laplace transforms (see Ref.\ \cite{hoyuelos} for further details).

For example, the probability of a micro-state $\omega$ with energy $U_\alpha$ and particle number $N_\alpha$ is
\begin{equation}\label{PomegaUN}
	P(\omega) = \frac{\exp[- \tfrac{1}{T} U_\alpha + \tfrac{\mu}{T}N_\alpha]}{\mathcal{Z}(\tfrac{1}{T},\tfrac{\mu}{T})}.
\end{equation}
Replacing $\mathcal{Z}(\tfrac{1}{T},\tfrac{\mu}{T}) = \exp(\tilde{S}[\tfrac{1}{T},\tfrac{\mu}{T}])$ and multiplying by the number of microstates $\Omega(U_\alpha,N_\alpha) = \exp(\tilde{S}(U_\alpha,N_\alpha))$ we have the probability $P_{(U_\alpha,N_\alpha)}$ as presented in Eq.\ \eqref{PX}.

\section{The probability ratio}
\label{s.pratio}

The ratio $P_B/P_A$ can be expressed as 
\begin{equation}
	\frac{P_B}{P_A} = \frac{P_{\vXl - \vd} P_{\vXr + \vd}}{P_{\vXl} P_{\vXr}}.
\end{equation}

Using Eq.\ \eqref{PX}, we obtain:
\begin{equation}\label{db2}
	\frac{P_B}{P_A} = \exp \left(\tS_{\vXr + \vd} - \tS_{\vXr} + \tS_{\vXl - \vd} - \tS_{\vXl} \right)
\end{equation}
where $\tS_{\vX_{\alpha}} = \tilde{S}(\vX_{\alpha})$ is introduced to simplify the notation.

Recall that $X_{\alpha i}$ (with $\alpha=$ L or R) represent the instantaneous values of extensive quantities in each cell, including fluctuations, while $X_i$ (without subindex $\alpha$) denotes the thermodynamic averages, which are the same for both cells. The intensive variables $Y_i = \frac{\partial S(\vX)}{\partial X_i}$, are determined by the reservoir.

We analyze equation \eqref{db2} by considering terms up to order $N^{-1}$. The average of $\delta_i$ in an infinitesimal time interval is $O(1)$; transition rates for $\delta_i$ of a larger order of magnitude are negligible. Defining the derivative operators $\partial_{\rm L} = \partial_{\vXl}$ and $\partial_{\rm R} = \partial_{\vXr}$, the  entropy difference in \eqref{db2} is expanded as:
\begin{equation}\label{difSex}
	\tS_{\vXr + \vd} - \tS_{\vXr} = \vd \cdot \partial_{\rm R} \tS_{\rm R} + \tfrac{1}{2} \vd\vd : \partial_{\rm R}^2 \tS_{\rm R},
\end{equation}
where $\tS_{\rm R} = \tS_{\vXr}$ and $\vd\vd : \partial_{\rm R}^2 \tS_{\rm R} = \sum_{i,j} \delta_i \delta_j \partial_{{\rm R}i}\partial_{{\rm R}j} \tS_{\rm R} $. The difference between SM and TH Massieu functions, representing the non-extensive part of $\tS_\alpha$, is defined as $\Delta S_\alpha = \tS_\alpha - S_\alpha$. This term is $O(\ln N)$, though for excess quantities, the difference is $O(N^0)$ \cite{hoyuelos}. Consequently, Eq.\ \eqref{difSex} can be rewritten in terms of the TH entropy $S_\alpha$ (without tilde), retaining terms up to $O(N^{-1})$:
\begin{equation}\label{difSex2}
	\tS_{\vXr + \vd} - \tS_{\vXr} = \underbrace{\vd\cdot\partial_{\rm R} S_{\rm R}}_{O(N^0)} + \underbrace{\vd\cdot \partial_{\rm R} \Delta S_{\rm R} + \tfrac{1}{2} \vd\vd : \partial_{\rm R}^2 S_{\rm R}}_{O(N^{-1})}.
\end{equation}
Following the same procedure for the left cell yields:
\begin{equation}\label{difSex2L}
	\tS_{\vXl} - \tS_{\vXl - \vd} = \vd\cdot\partial_{\rm L} S_{\rm L} + \vd\cdot \partial_{\rm L} \Delta S_{\rm L} - \tfrac{1}{2} \vd\vd : \partial_{\rm L}^2 S_{\rm L}.
\end{equation}

By defining the operator $\mathcal{D} = \vd\cdot(\partial_{\rm R} - \partial_{\rm L})$, the combined entropy difference becomes:
\begin{align}
\tS_{\vXr + \vd} & - \tS_{\vXr} + \tS_{\vXl - \vd} - \tS_{\vXl} =\nonumber\\ 
& \mathcal{D} (S_{\rm L} + S_{\rm R}) + \mathcal{D} (\Delta S_{\rm L} + \Delta S_{\rm R}) + \frac{1}{2} \mathcal{D}^2 (S_{\rm L} + S_{\rm R}),
\end{align}
where the identity $\partial_{\rm L}\partial_{\rm R}(S_{\rm L} + S_{\rm R})=0$ was used. Substituting these into \eqref{db2}, the probability ratio is:
\begin{align}\label{pratio}
	\frac{P_B}{P_A} &= \exp \left[\mathcal{D} (S_{\rm L} + S_{\rm R}) + \mathcal{D} (\Delta S_{\rm L} + \Delta S_{\rm R}) \right. \nonumber\\
	&\left. \quad + \tfrac{1}{2} \mathcal{D}^2 (S_{\rm L} + S_{\rm R}) \right] \nonumber\\
	&= e^{\mathcal{D} (S_{\rm L} + S_{\rm R})}\,\left[ 1 + \mathcal{D} (\Delta S_{\rm L} + \Delta S_{\rm R}) \right. \nonumber\\
	& \quad \left. + \tfrac{1}{2} \mathcal{D}^2 (S_{\rm L} + S_{\rm R}) \right],
\end{align}
where the exponential of the $O(N^{-1})$ terms has been linearized in the final step.

\section{Transition rate}
\label{s.tr}

We explicitly express the dependence of the transition rate on the variables of the initial state and the transported quantity: $W_{AB}(\vXl,\vXr,\vd)$. Assuming that the system is isotropic, we have the following relationship for $W_{BA}$:
\begin{equation}
	W_{BA} = W_{AB}(\vXl - \vd,\vXr + \vd,-\vd)
\end{equation}
where the minus sign in the last argument ($-\vd$) indicates  transport from R to L. Using a Taylor expansion around $(\vXl,\vXr)$, we obtain:
\begin{align}
	W_{BA} = & W_{AB}(\vXl,\vXr,-\vd) + \vd\cdot\partial_{\rm R} W_{AB}(\vXl,\vXr,-\vd) \nonumber\\
	&  - \vd\cdot\partial_{\rm L}W_{AB}(\vXl,\vXr,-\vd)
\end{align}

It is convenient to adopt the following compact notation: 
\begin{align}
	W_{AB} &= W_{\vd},\label{notWab} \\
	W_{BA} &= W_{-\vd} + \mathcal{D} W_{-\vd}.\label{notWba}
\end{align}

With these definitions, the detailed balance condition \eqref{detbal} yields:
\begin{equation}
	W_{\vd} = W_{-\vd}\,(1 + \mathcal{D} \ln W_{-\vd}) \frac{P_B}{P_A}.
\end{equation}
Using Eq.\ \eqref{pratio} for $P_B/P_A$ and neglecting terms $O(N^{-2})$, we have:
\begin{align}
W_{\vd}& = W_{-\vd}\,e^{\mathcal{D} (S_{\rm L} + S_{\rm R})} \, \left[ 1 + \mathcal{D} \ln W_{-\vd} \right. \nonumber\\
& \quad \left. + \mathcal{D} (\Delta S_{\rm L} + \Delta S_{\rm R})  + \tfrac{1}{2} \mathcal{D}^2 (S_{\rm L} + S_{\rm R})\right].
\end{align}

The next step is to separate orders $N^0$ and $N^{-1}$. At $O(N^0)$ we obtain:
\begin{equation}\label{ON0}
	W_{\vd} = W_{-\vd}\,e^{\mathcal{D} (S_{\rm L} + S_{\rm R})},
\end{equation}
and at $O(N^{-1})$:
\begin{equation}\label{ONm12}
\mathcal{D} \ln W_{-\vd} =  - \mathcal{D} (\Delta S_{\rm L} + \Delta S_{\rm R}) - \tfrac{1}{2} \mathcal{D}^2 (S_{\rm L} + S_{\rm R}).
\end{equation}
To be rigorous, a constant has to be introduced within $\ln W_{-\vd}$ to ensure the argument is dimensionless, though it is omitted here as it does not affect the subsequent calculations.
The solution is:
\begin{equation}\label{wrl}
	\ln (W_{-\vd}) = - \Delta S_{\rm L} - \Delta S_{\rm R} - \tfrac{1}{2} \vd\cdot\Delta \vY + f_{\rm LR},
\end{equation}
where we define $\Delta \vY = \partial_{\rm R} S_{\rm R} - \partial_{\rm L} S_{\rm L}$, and $f_{\rm LR}$ satisfies $\mathcal{D} f_{\rm LR}=0$. This condition is fulfilled if $f_{\rm LR}$ is a function of the average state $\vXm=(\vXl + \vXr)/2$. 

We can approximate the previous expression as:
\begin{equation}
	- \Delta S_{\rm L} - \Delta S_{\rm R} + f_{\rm LR} = - 2\Delta S + f + \vec{h}\cdot \Delta\vY,
\end{equation}
where, in the right hand side, we have the same terms but evaluated at $\vXl=\vXr=\vX$ (quantities without subindex L or R) plus a correction that is proportional to $\Delta\vX = \vXr - \vXl$ and that was rewritten as proportional to $\Delta \vY$ (using the relation $\Delta\vX = \frac{\partial\vX}{\partial\vY}\cdot \Delta\vY$); $\vec{h}$ is the proportionality vector (as shown later, the term with $\vec{h}$ does not contribute to the currents of conserved quantities). Thus,
\begin{equation}\label{e.wd0}
	W_{\vd} = e^{- 2\Delta S + (\frac{1}{2} \vd + \vec{h})\cdot\Delta \vY + f}.
\end{equation}

Next, we use an ideal reference system, for which transport coefficients are known, to eliminate the unknown function $f$.  In the ideal system, interactions among constituent elements are neglected, simplifying the determination of the equation of state. We assume that interactions can be gradually reduced such that Eq.\ \eqref{e.wd0} is continuously transformed into the transition rate equation for the ideal system. To avoid divergences in this procedure, the real and ideal systems must remain in the same phase, as $\Delta S$ depends on thermodynamic fluctuations.

Evaluating the ideal system at $\vXl=\vXr=\vX$ (such that $\Delta\vY=0$), we have:
\begin{equation}\label{e.wdid}
	\bar{W}_{\vd}^\text{id} = e^{-2\, \Delta S^\text{id} + f},
\end{equation}
where the bar in $\bar{W}_{\vd}^\text{id}$ denotes that the transition rate is evaluated at the thermodynamic state $\vX$. Combining Eqs.\ \eqref{e.wd0} and \eqref{e.wdid}, we obtain,
\begin{equation}\label{e.wd}
	W_{\vd} = \bar{W}_{\vd}^\text{id} e^{-2\, \Delta S^\text{e} + (\frac{1}{2} \vd + \vec{h})\cdot\Delta \vY}
\end{equation}
where $\Delta\Se = \Delta S - \Delta S^\text{id}$ is the excess part of $\Delta S$. The difference $\Delta\Se = \tSe - \Se$ can be expressed in terms of the Hessian matrix $H_{i,j} = \frac{\partial^2 S(\vX)}{\partial X_i \partial X_j}$ (for $i,j = 1\cdots n$) as (see \cite{hoyuelos}):
\begin{equation}\label{deltaS}
	\Delta \Se = \frac{1}{2}  \ln \frac{\det(H)}{\det(H_\text{id})},
\end{equation}
where $H_\text{id}$ is the Hessian of $S^\text{id}$. The  covariance matrix $C = \langle (\vX_\alpha - \vX)(\vX_\alpha-\vX)^T\rangle$ is equal to $-H^{-1}$.
In thermodynamic equilibrium ($\vXl=\vXr=\vX$, $\Delta\vY=0$), from \eqref{e.wd}, the transition rate becomes:
\begin{equation}\label{e.wdeq}
	\bar{W}_{\vd} = \bar{W}_{\vd}^\text{id} e^{-2\, \Delta S^\text{e}} = \bar{W}_{\vd}^\text{id} \frac{\det C}{\det C_\text{id}}.
\end{equation}

If only a subset of $s$ extensive variables, $\vX_\alpha^s = (X_{\alpha,1}\cdots X_{\alpha,s})$, undergoes transitions between cells, the probability of this subset is: 
\begin{equation}
	P(\vX_\alpha^s) = e^{-\tilde{S}[\vY] + \tilde{S}[\vX_\alpha^s] - Y_1 \hat{X}_1 \dots - Y_s \hat{X}_s},
\end{equation}
where $\tilde{S}[\vX_\alpha^s] = \tS[X_{\alpha,1}\dots X_{\alpha,s}, Y_{s+1} \dots Y_{n}]$ is the corresponding Massieu function; see Eq.\ (17) in Ref.\ \cite{hoyuelos}. Following the same procedure, Eq. \eqref{e.wd} is recovered, where the non-extensive term of the excess part of the Massieu function is given by
\begin{equation}\label{deltaSHs}
	\Delta \Se[\vX^s] = \frac{1}{2}  \ln \frac{\det(H^s)}{\det(H_\text{id}^s)},
\end{equation}
with $H^s_{i,j} = \frac{\partial^2 S[\vX^s]}{\partial X_i \partial X_j}$, with $i,j = 1\cdots s$.

\section{Transport coefficients}
\label{s.transp}

To calculate transport coefficients, we consider a small departure from equilibrium driven by thermodynamic forces, which are represented by small gradients of intensive variables. 

In the previous section, Eq.\ \eqref{wrl} defined the transition rate, $W_{-\vd}$, for the state $(\vXr,\vXl)$. This state may differ slightly from thermodynamic equilibrium due to fluctuations. (Recall that in this notation, quantities with the subscript $\alpha=$ R or L represent the macro-state for a particular micro-state, while those without subscript represent thermodynamic averages). The situation we analyze now is distinct: the departure from equilibrium is produced not by fluctuations, but by applied thermodynamic forces. Nevertheless, because the departure is assumed to be very small in both cases, we can assume that Eq.\ \eqref{wrl} remains valid. In this context, we interpret the difference $\Delta \vY$ as an applied force rather than a fluctuation. 

The vector $\vX$ represents the fluctuating extensive quantities; the transport of $\vd$ from cell L to cell R occurs with a transition rate $W_{\vd}$, given by \eqref{e.wd}. The net current of $\vX$ from L to R is:
\begin{align}\label{current}
	\vec{J} &=  \frac{1}{a^2} \sum_{\vd} \vd\, W_{\vd}  \nonumber\\
	& = \frac{1}{a^2} e^{-2\, \Delta S^\text{e}}  \sum_{\vd} \vd\,\bar{W}_{\vd}^\text{id} e^{(\frac{1}{2} \vd + \vec{h})\cdot\Delta \vY} \nonumber\\
	& = \frac{1}{a^2} e^{-2\, \Delta S^\text{e}}  \sum_{\vd} \vd\,\bar{W}_{\vd}^\text{id} \left[1 + (\tfrac{1}{2} \vd + \vec{h})\cdot\Delta \vY\right]
	\nonumber\\
	& = \frac{1}{2 a^2} e^{-2\, \Delta S^\text{e}}  \sum_{\vd} \bar{W}_{\vd}^\text{id}\, \vd\vd^T\cdot\Delta \vY
\end{align}
where $a^2$ is the contact area between cells L and R. We  assumed that $\Delta \vY$ is small and utilized the fact that $\bar{W}_{\vd}^\text{id}$ is an even function of $\vd$. The sum is performed over all possible values of the components of $\vd$; depending on the system, this sum may be replaced by an integral. Thermodynamic forces are defined as the negative gradient of the intensive variables, that we approximate as the finite difference ratio:
\begin{equation}\label{force}
	\vF = -\frac{\Delta \vY}{a},
\end{equation}
For example, for a system described by internal energy $U$ and particle number $N$, the intensive parameters are $\vY = (\frac{1}{T},-\frac{\mu}{T})$. The resulting forces are $\vec{F} = \frac{1}{a}\left(\frac{1}{T^2}\Delta T,\Delta(\frac{\mu}{T})\right)$.

As long as the force is sufficiently small, a linear relation exists between the current and the force:
\begin{equation}\label{linear}
	\vJ = -L \cdot \vF,
\end{equation}
where $L$ is the Onsager matrix of phenomenological coefficients. 

We can obtain matrix $L$ from Eq.\ \eqref{current}:
\begin{equation}\label{current2}
	\vec{J} = -\underbrace{\frac{1}{2 a} \frac{\det(H_\text{id})}{\det(H)}\,  \left( \sum_{\vd} \bar{W}_{\vd}^\text{id}\, \vd\vd^T \right)}_{L} \cdot \vF,
\end{equation}
where Eq.\ \eqref{deltaS} was employed. The Onsager matrix is thus:
\begin{equation}\label{L}
	L =  \frac{\det(H_\text{id})}{\det(H)}\, L_{\rm id},
\end{equation}
where
\begin{equation}\label{Lid}
	L_{\rm id} = \frac{1}{2a} \sum_{\vd} \bar{W}_{\vd}^{\rm id}\, \vd\vd^T,
\end{equation}
is the matrix for the ideal case, which satisfies the symmetry requirements of the Onsager relations. Using the covariance matrix $C$, we can also write
\begin{equation}\label{LC}
	L =  \frac{\det(C)}{\det(C_\text{id})}\, L_{\rm id}.
\end{equation}
Consequently, if the Onsager matrix for the ideal case ($L_{\rm id}$) is known,  Eq.\ \eqref{LC} provides the Onsager matrix $L$ for the interacting system, provided the equation of state is available to calculate $C$ or $H$.

\subsection{Interpretation from thermodynamic geometry}

Some physical intuition behind Eq.\ \eqref{L} can be obtained from thermodynamic geometry. The geometric foundation rests on the theory of dually flat thermodynamic manifolds, a framework rooted in Information Geometry \cite{amari,amari2} and extended to the geometric theory of heat in Ref.\ \cite{barbaresco}; see also \cite{nielsen}. In this context, the space of equilibrium states is viewed as a landscape where the geometry is defined by the Legendre duality between extensive variables $\vX$ and intensive parameters $\vY$. The term 'dually flat' signifies that the manifold's structure is symmetrically linked through these dual representations: while the entropy $S(\vX)$ generates the geometry in the extensive space, its total Legendre transform, the Massieu function $S[\vY]$, defines the same manifold in the intensive space.

The Hessian matrix $H$ of the entropy defines the metric tensor of the state space \cite{ruppeiner}. The square root of the determinant of the metric tensor acts as a measure of information density, telling us how densely the underlying microscopic states are packed within a given macroscopic state. Thus, the transport matrix $L$ is given by a ratio of these densities, comparing how the landscape of a real system, distorted by interparticle interactions, differs from the landscape of an ideal reference.

\subsection{Force $\vF' = A\cdot\vF$.}

A generalization, considered here, is necessary when forces are defined in specific ways. For example, the force driving a particle current is often taken as the concentration gradient rather than the gradient of the chemical potential over temperature. Let us assume that the new forces, $\vF'$, are related to $\vF$ through $\vF' = A\cdot\vF$, with $A$ an invertible matrix. 
The current is
\begin{equation}
	\vJ = -L\cdot \vF = -L\cdot A^{-1}\cdot A\cdot\vF = -L'\cdot \vF',
\end{equation}
where 
\begin{equation}
	L' = L\cdot A^{-1}
\end{equation}
is the new Onsager matrix, not necessarily symmetric. Using that $L_\text{id} = L_\text{id}' \cdot A_\text{id}$ and Eq.\ \eqref{LC}, we have
\begin{equation} \label{Lp0}
	L' = \frac{\det(C)}{\det(C_\text{id})} \,L_{\rm id}'\cdot A_{\rm id}\cdot A^{-1}.
\end{equation}

\section{Diffusivity}
\label{s.diffusion}

Consider particles diffusing on a lattice. We analyze the system as a binary mixture of species $A$ and $B$ within a single lattice site coupled to a reservoir at temperature $T$ and chemical potentials $\mu_A$ and $\mu_B$. This approach is applicable to diffusion in solids or on surfaces where an underlying lattice is present. Here, the model site represents a coarse-grained cell that encompasses multiple crystalline sites and contains many particles. The system is a binary mixture of $N_A$ particles (species $A$) and $N_B$ particles (species $B$), with a total particle count of $N = N_A + N_B$. At constant $T$, the Massieu function  $\Phi = S[T,N_A,N_B] = S - \frac{1}{T}U$ (equal to minus the Helmholtz free energy over the temperature) obeys:
\begin{equation}\label{dPhi0}
    d\Phi = -\frac{\mu_A}{T} dN_A - \frac{\mu_B}{T} dN_B.
\end{equation}

\subsection{Diffusion in solids --- Darken model}

In solids, diffusion of similar species primarily occurs through vacancies via a process known as substitutional diffusion. In the Darken model for substitutional diffusion  \cite{darken}, it is assumed that the concentration of vacancies is at equilibrium and that the total number of particles $N= N_A + N_B$ is constant. Therefore, Eq.\ \eqref{dPhi0} can be rewritten as
\begin{equation}
    d\Phi = -\frac{\mu_{AB}}{T} dN_A
\end{equation}
with $\mu_{AB} = \mu_A - \mu_B$. From the Gibbs-Duhem equation, it can be shown that the thermodynamic factor $\Gamma_{AB} = \frac{\partial\mu_A/T}{\partial\log N_A} = \frac{\partial\mu_B/T}{\partial\log N_B}$ is the same for the species $A$ or $B$. $J_A$ is the current of $A$ particles, the corresponding force is the gradient of $\mu_{AB}/T$, and these are related through the coefficient $L_A$:
\begin{equation}
    J_A = -L_A\, \nabla \left(\frac{\mu_{AB}}{T}\right) = -\frac{L_A}{C_A} \,\nabla N_A,
\end{equation}
where $C_A = \frac{\partial N_A}{\partial \mu_{AB}/T}=N_A N_B/(N \Gamma_{AB})$ represents the squared fluctuations of $N_A$. Then, the intrinsic diffusion coefficient for species $A$ is
\begin{equation} \label{e.DA}
    D_A = \frac{L_A}{C_A} = \frac{L_A^\text{id}}{C_A^\text{id}}=D_A^\text{id}
\end{equation}
where Eq.\ \eqref{LC} was used, from which we have $L_A = L_A^\text{id}\, C_A/C_A^\text{id}$ (also taking into account that, in this case, $C_A$ is a scalar). An equivalent equation is obtained for species $B$.

We can obtain the tracer diffusion coefficient of a tagged particle $A$, $D_A^*$. The particle performs a random walk with average transition rate $\bar{W}_1$ and jump size $a$. Using the continuous limit of a random walk, the tracer diffusion coefficient is 
\begin{equation}
D_A^* = a^2 \bar{W}_1 = a^2 \bar{W}_1^\text{id} \frac{C_A}{C_A^{\rm id}},
\end{equation}
where Eq.\ \eqref{e.wdeq} with $\delta=1$ was used. In the ideal case, when interactions are absent, $\Gamma_{AB}^\text{id}=1$, and $C_A^\text{id}=N_A N_B/N$; also, $a^2 \bar{W}_1^\text{id} = D_A^\text{id}$, that is, the tracer diffusivity is equal to the intrinsic diffusivity in the ideal case. Then, using \eqref{e.DA}, we have
\begin{equation}\label{e.dark}
    D_A^* = \frac{D_A}{\Gamma_{AB}}.
\end{equation}
This result reproduces one of the Darken equations for substitutional diffusion in binary alloys and verifies the validity of the present procedure. Darken's approach does not take into account off-diagonal elements in the Onsager matrix; see, for example, \cite{mehrer,shewmon}. Although the off-diagonal elements can be ignored in many situations, Binder \cite{binder} pointed out that they are not always negligible with respect to the diagonal ones. Zhang \textit{et al.}\ \cite{zhang}, using molecular dynamics simulations of a fluid mixture of Ar and Kr, showed that the off-diagonal contribution (related to what they call distinct diffusivity or $D_d$) is close to zero in the bulk, but becomes more relevant for diffusion in nanopores. Hartmann \textit{et al.}\ \cite{hartmann} numerically analyzed diffusion processes in 3D of two species to conclude that an accurate physical description requires the inclusion of the off-diagonal terms in the Onsager matrix. 

In the next section, a different model, appropriate for surface diffusion, is considered, where off-diagonal elements are present.

\subsection{Diffusion on surfaces}
\label{s.surf}

Let us consider again the binary mixture described by Eq.\ \eqref{dPhi0}. The primary difference compared to the Darken model is that the total particle number, $N=N_A+N_B$, is no longer constant. 
For simplicity, we assume all particles are physically identical, meaning species $A$ and $B$ are equivalent except for their identifying tags. Consequently, the chemical potentials are $\mu_A = \mu^0 + T\log N_A + \mu^\text{e}$ and $\mu_B = \mu^0 + T\log N_B + \mu^\text{e}$, where $\mu^0$ is a reference value and $\mu^\text{e}$ is the excess chemical potential (a function of $N$). Both $\mu^0$ and  $\mu^\text{e}$ are identical for both species. While we consider a two-dimensional system here, these calculations can be generalized to other dimensions.

By defining the difference between tagged and normal particles as $N_D = N_A - N_B$, we can rewrite the differential of the Massieu function from \eqref{dPhi0} in terms of $N$ and $N_D$:
\begin{equation}\label{dS2}
	d\Phi =  - \frac{\mu_N}{T} dN  - \frac{\mu_D}{T} dN_D
\end{equation}
with
\begin{align}
	\frac{\mu_N}{T} &= \frac{\mu_A+\mu_B}{2T} = \frac{\mu}{T} + \frac{1}{2}\ln\left( \frac{N^2 - N_D^2}{4 N^2} \right), \\
	\frac{\mu_D}{T} &= \frac{\mu_A-\mu_B}{2T} = \frac{1}{2} \ln\left( \frac{N + N_D}{N - N_D} \right),
\end{align}
where $\mu = \mu^0 + T\log N + \mu^\text{e}$. 
The extensive variables are $\vX = (N,N_D)$ and the intensive ones are $\vY = (-\frac{\mu_N}{T},-\frac{\mu_D}{T})$. The corresponding covariance matrix is
\begin{align}\label{fluctmatrix}
	C &= \left( \begin{array}{cc}
		\left. \frac{\partial N}{\partial \frac{\mu_N}{T}}\right|_\frac{\mu_D}{T} & \left. \frac{\partial N}{\partial \frac{\mu_D}{T}}\right|_\frac{\mu_N}{T} \\
		\left. \frac{\partial N_D}{\partial \frac{\mu_N}{T}}\right|_\frac{\mu_D}{T} & \left. \frac{\partial N_D}{\partial \frac{\mu_D}{T}}\right|_\frac{\mu_N}{T} 
		\end{array} \right) \nonumber\\
	&= N\left( \begin{array}{cc}
		1/\Gamma & x/\Gamma \\
		x/\Gamma & 1 - (1-\frac{1}{\Gamma})x^2
	\end{array} \right),
\end{align}
where $\Gamma = N\left.\frac{\partial \mu/T}{\partial N} \right|_T$ is the thermodynamic factor, and $x=N_D/N$ is the fraction of the particle difference.
In the ideal case ($\Gamma = 1$), the covariance matrix $C_\text{id}$ is equivalent to $C$ with $\Gamma=1$; $x$ remains unchanged since, by definition, the ideal system has the same extensive variables $\vX$ as the original system. The ratio of the determinants is
\begin{equation}\label{determ}
	\frac{\det(C)}{\det(C_\text{id})} = \frac{1}{\Gamma}.
\end{equation}

We are interested in the Onsager matrix that connects the collective current $j$ of $N$ and the diffusion current $j_D$ of $N_D$ to the forces. 
The system is isotropic and the currents $j$ and $j_D$ represent components in an arbitrary direction (see the scheme of Fig.\ \ref{f.scheme}).
When interactions are neglected (the ideal case), and considering concentration gradients as the driving forces, we have:
\begin{equation}
		\left(\begin{array}{c}
		j_\text{id} \\
		j_{D,\text{id}}
	\end{array}\right) = - L_\text{id}' \cdot \left(\begin{array}{c}
\Delta \rho/a \\
\Delta \rho_D/a
\end{array}\right)
\end{equation}
where $\rho=N/a^2$ and $\rho_D=N_D/a^2$, with $a^2$ being the lattice site area. Here, the diffusion coefficient $D_\text{id}$ is the identical for both $N$ and $N_D$:  
\begin{equation}\label{Lidp}
	L_\text{id}' = \left(\begin{array}{cc}
		D_\text{id} & 0 \\
		0 & D_\text{id}
	\end{array}\right).
\end{equation}
Note that $L_\text{id}$ is not equivalent to $L_\text{id}'$, as the forces that we have to consider are the gradients of $\vY_\text{id}$ rather than the concentration gradients: 
\begin{equation}
	\left(\begin{array}{c}
		j_\text{id} \\
		j_{D,\text{id}}
	\end{array}\right) = -L_\text{id} \cdot \left(\begin{array}{c}
	\Delta (-\frac{\mu_N}{T})/a \\
	\Delta (-\frac{\mu_D}{T})/a
\end{array}\right)
\end{equation}
These forces are related via the transformation:
\begin{equation}\label{transfA}
	\left(\begin{array}{c}
		\Delta \rho \\
		\Delta \rho_D
	\end{array}\right) = A\cdot \left(\begin{array}{c}
	\Delta (-\frac{\mu_N}{T}) \\
	\Delta (-\frac{\mu_D}{T})
\end{array}\right),
\end{equation}
where $A= - \frac{1}{V}C$. 
Using \eqref{Lp0}, we find:
\begin{equation}\label{Lp}
	L'= \frac{1}{\Gamma} L_\text{id}'\cdot C_\text{id} \cdot C^{-1} = D_\text{id}\left( \begin{array}{cc}
		1 & 0 \\
		(1-\frac{1}{\Gamma})x & \frac{1}{\Gamma}
	\end{array} \right).
\end{equation}
The resulting currents are:
\begin{align}
	\vec{j} &=  -D_\text{id}\, \nabla \rho \label{e.jN}\\
	\vec{j}_D &=  -D_\text{id}(1-\tfrac{1}{\Gamma}) x\, \nabla \rho - D_\text{id}\tfrac{1}{\Gamma}\, \nabla\rho_D, \label{e.jD}
\end{align}
where, invoking the isotropy of the system, the results were generalized to any direction, and, taking the continuous limit, the finite differences were replaced by gradients. The specific currents for species $A$ and $B$ can then be derived from \eqref{e.jN} and \eqref{e.jD}:
\begin{align}
	\vec{j}_A &=  -\frac{D_\text{id}}{\rho}\, \left( (\rho_A + \rho_B/\Gamma)\, \nabla \rho_A + (1-1/\Gamma)\rho_A \,\nabla \rho_B \right) \label{e.jA}\\
	\vec{j}_B &=  -\frac{D_\text{id}}{\rho}\, \left( (1-1/\Gamma)\rho_B \, \nabla \rho_A +  (\rho_B + \rho_A/\Gamma) \,\nabla \rho_B \right), \label{e.jB}
\end{align}
where $\rho_A = N_A/a^2$ and $\rho_B = N_B/a^2$.

Models for surface diffusion often utilize Langmuir particles characterized by an exclusion process \cite{liggett}, which implies hard-core pair interactions at the microscopic level. Within the coarse-grained approximation employed here, these microscopic interactions manifest as a maximum particle occupancy $N_m$ per lattice site, modeled as a soft-core interaction. The corresponding thermodynamic factor is,
\begin{equation}\label{e.gsc}
	\Gamma = \frac{1}{1 - \rho/\rho_m},
\end{equation}
where $\rho_m= N_m/a^2$ [see Eq.\ (33) in \cite{dimuro} or Eq.\ (2.106) in \cite{gomer}]. The $A$ and $B$ particle currents become:
\begin{align}
	\vec{j}_A &=  -\frac{D_\text{id}}{\rho_m}\, \left[ (\rho_m - \rho_B)\, \nabla \rho_A + \rho_A \,\nabla \rho_B \right] \\
	\vec{j}_B &=  -\frac{D_\text{id}}{\rho_m}\, \left[ \rho_B \, \nabla \rho_A +  (\rho_m - \rho_A) \,\nabla \rho_B \right]. 
\end{align}
Expressing these in terms of the fractional coverages $\theta_A = \rho_A/\rho_m$ and $\theta_B = \rho_B/\rho_m$, and defining $\theta=\theta_A + \theta_B$ and $\theta_D = \theta_A - \theta_B$, the corresponding currents for coverages are:
\begin{align}
	\vec{j} &=  -D_\text{id}\, \nabla \theta \label{e.jtheta}\\
	\vec{j}_D &=  -D_\text{id}(1-\tfrac{1}{\Gamma}) x\, \nabla \theta - D_\text{id}\tfrac{1}{\Gamma}\, \nabla\theta_D, \label{e.jthetaD}
\end{align}
or,
\begin{align}
	\vec{j}_A &=  -D_\text{id} \left[ (1 - \theta_B)\, \nabla \theta_A + \theta_A \,\nabla \theta_B \right], \label{e.jAsc}\\
	\vec{j}_B &=  -D_\text{id}\, \left[ \theta_B \, \nabla \theta_A +  (1 - \theta_A) \nabla \theta_B \right], \label{e.jBsc}
\end{align}
where the replacement $\vec{j} \rightarrow \rho_m \vec{j}$ has been performed in \eqref{e.jN}, and similarly for the other currents, to represent currents of coverages instead of currents of particles.
These equations reproduce the Zhdanov model for the diffusion of coadsorbed Langmuir particles on surfaces; see Eqs.\ (7) and (8) in \cite{zhdanov0}, or Eq.\ (3) in \cite{zhdanov}. However, it should be noted that dynamic correlations produced by, for example, microscopic exchange jumps (where two particles swap places, $AB \leftrightarrows  BA$) and pair jumps (where two adjacent particles move together to two vacant sites) are not properly represented by this model \cite{zhdanov}. In this case, the coarse-grained approximation leading to the soft-core interaction and the thermodynamic factor of Eq.\ \eqref{e.gsc} should be revised.

\section{Numerical results}
\label{s.numerical}

We present numerical simulation results that verify the analytical findings of Sec.\ \ref{s.surf}. 
From Eq.\ \eqref{e.jN}, the collective diffusion coefficient is given by $D_c = D_\text{id}$, while from \eqref{e.jD}, the self-diffusion coefficient is $D^* = D_\text{id}/\Gamma$. Note that $D^*$ is equivalent to the tracer diffusion coefficient when the tagged particle interacts with particles of the same kind. Then,
\begin{equation}
    D^* = \frac{D_c}{\Gamma},
\end{equation}
a relationship equivalent to the Darken equation \eqref{e.dark}. This relation was numerically verified in Ref.\ \cite{dimuro} for different interaction types. In this section, we focus on verifying the off-diagonal terms of the diffusion matrix. We employ a soft-core interaction, representing a coarse-grained approximation of microscopic hard-core interactions, as it is particularly suited for modeling surface diffusion \cite{ala,gomer,zhdanov0,zhdanov}. Additionally, we examine several phenomenological models where the excess chemical potential has a linear, quadratic or cubic dependence on the concentration. These models serve to provide a broader data set, supporting the generality of our analytical results.

\subsection{Simulation details}

The simulations consist of a diffusive Monte Carlo (MC) process on a 1D lattice gas with $\mathcal{L}$ sites. The system contains $\mathcal{N}=\mathcal{N}_A+\mathcal{N}_B$ particles of types $A$ and $B$, each site $i$ containing a number $N_{A,i}$ and $N_{B,i}$, such that $\mathcal{N}_A=\sum_i N_{A,i}$ and $\mathcal{N}_B=\sum_i N_{B,i}$. Particles jump between neighboring sites with a rate that generally depends on the number of particles at the origin and destination sites. For instance, for particles that jump from left to right the rate is $W(N_\text{L},N_\text{R})$, where $N_\text{L}$ and $N_\text{R}$ are the total number of particles in the right and left sites, respectively. Likewise, the jump rate for particles that jump from right to left is $W(N_\text{R},N_\text{L})$. The coverage of type-$A$ particles at site i is $\theta_{A,i} = N_{A,i}/N_m$,  were $N_m$ plays the role of the volume of a cell. Equivalent expressions hold for type-$B$ particles. The average coverage of the system is $\bar{\theta}=\bar{\theta}_A+\bar{\theta}_B$, where $\bar{\theta}_A = \sum_{i=0}^{\mathcal{L}} \theta_{A,i}/(\mathcal{L}+1)$.

\begin{figure}
    \centering
    \includegraphics[width=1\linewidth]{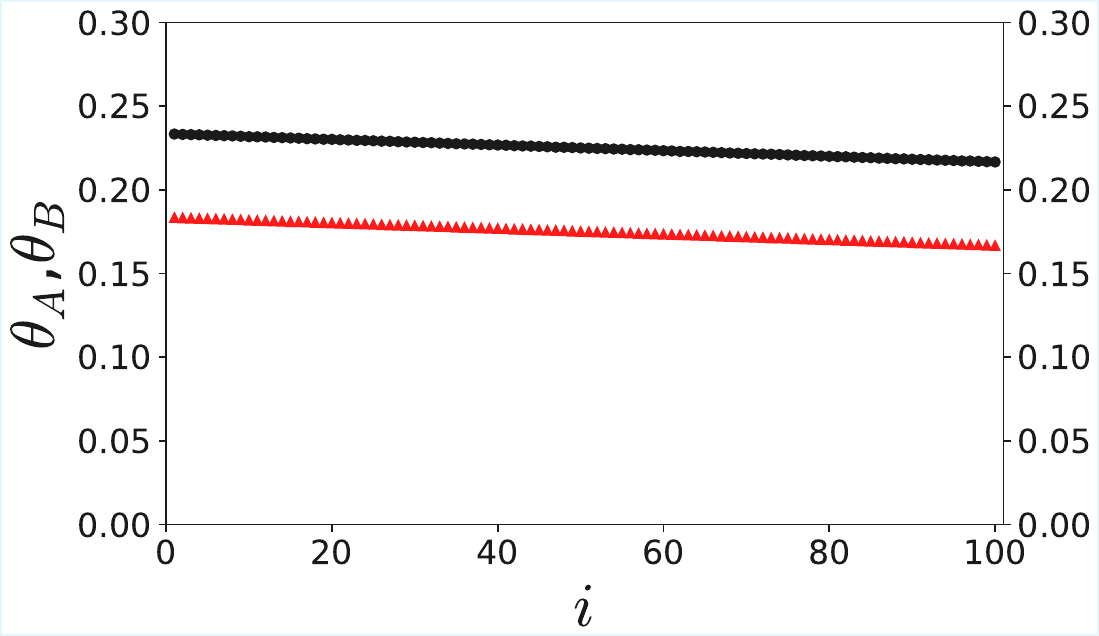}
    \caption{Coverage profile of a 1D lattice of size $\mathcal{L}$ at the steady state, of particles of type $A$ (black dots) and $B$ (red diamonds). At $i=0$ the type-$A$ and $B$ particle coverage is fixed at parameters $c_A$ and $c_B$, respectively, while for $i=\mathcal{L}$ is fixed at $d_A$ and $d_B$. The slopes are equal, given by Eq.~\eqref{e.grad}. In this particular case, we have $N_m=600$ and a mean coverage $\bar{\theta}=0.4$. The results were obtained from Monte Carlo simulations.}
    \label{fig:Histo}
\end{figure}

In order to keep a steady current, a density gradient is imposed, which is equal for both types of particles. For this purpose, the type $A$ and $B$ coverarages at site $i=0$ are fixed at parameters $c_A$ and $c_B$ respectively. Likewise, at site $i=\mathcal{L}$ the coverages of type $A$ and $B$ are fixed at parameters $d_A$ and $d_B$, respectively. At the steady state, the gradients of each species are constant and equal:   
\begin{equation}\label{e.grad}
\partial_x \theta_A=\frac{d_A-c_A}{a\mathcal{L}}=\frac{d_B-c_B}{a\mathcal{L}}=\partial_x \theta_B,
\end{equation}
where $\partial_x$ is the gradient along the horizontal axis, with $x = a\,i$, and $a$ is the lattice spacing.
Since both gradients are equal then $\partial_x\theta_D=0$; this condition is chosen in order to separately analyze the effect of the total coverage gradient in the relative or diffusion current $j_D$; see Eq.\ \eqref{e.jthetaD}. In Fig.~\ref{fig:Histo} we show an example of the coverage profile.

The dynamics evolve as follows. At each Monte Carlo step, $\mathcal{N}$ particles are selected at random to perform a jump. A particle (of type $A$ or $B$) located at site $i$ hops to the right with probability $p_\text{R}=W_{N_i,N_{i+1}}$ or to the left with probability $p_\text{L}=W_{N_i,N_{i-1}}$, where the argument order indicates that the jump is from $i$ to $i+1$ and from $i$ to $i-1$, respectively. Since particles $A$ and $B$ have the same physical properties, the transition rate is the same for both kinds. The instantaneous transition rate is obtained from \eqref{wrl} considering $\Delta Y = -(\mu_\text{R} - \mu_\text{L})/T$ and $\delta=1$ (the probability that more than one particle jumps in a short time interval is neglected):
\begin{equation}
    W_{N_\text{L},N_\text{R}} = W_{N_\text{L},N_\text{R}}^\text{id} e^{-\Delta S^\text{e}_\text{L} -\Delta S^\text{e}_\text{R} -(\mu^\text{e}_\text{R} - \mu^\text{e}_\text{L})/T},
\end{equation}
and
\begin{equation}
    W^\text{id}_{N_\text{L},N_\text{R}} = e^{-\Delta S^\text{id}_\text{L} -\Delta S^\text{id}_\text{R} -(\mu^\text{id}_\text{R} - \mu^\text{id}_\text{L})/T + f_\text{LR}}.
\end{equation}
In this simplified model it is assumed that, in the limit of small concentration (the ideal case), the transition rate is constant: $W^\text{id}_{N_\text{L},N_\text{R}} = \nu$. The excess part of the non-extensive term is $\Delta S^\text{e}_\alpha = \frac{1}{2} \ln \Gamma_\alpha$. Then,
\begin{equation} \label{e.WNN}
    W_{N_\text{L},N_\text{R}} = \nu \frac{e^{-(\mu^\text{e}_\text{R} - \mu^\text{e}_\text{L})/T}}{(\Gamma_\text{R} \Gamma_\text{L})^{1/2}},
\end{equation}
[see Eq.\ (23) in Ref.\ \cite{dimuro}].

In order to fix the density gradient, we apply the following rules for type $A$ particles at the borders of the lattice; equivalent rules also apply to type $B$ particles:

\medskip
\noindent (1) If the $A$ particle is located at $i=0$, and if jumps to the right, a particle of the same type is created at $i=0$. If instead, a hop to the left is attempted, the move is rejected.

\medskip
\noindent (2) If the selected  particle of type $A$ is at site $i=1$ and hops to the left, the particle vanishes. 

\medskip
\noindent (3) If the  particle of type $A$ is at site $i=\mathcal{L}-1$ and hops to the right, the particle also vanishes. 

\medskip
\noindent (4) If the location is $i=\mathcal{L}$, and if the particle of type $A$ hops to the left, another particle of the same type is created at $i=\mathcal{L}$. 
If the rightward move is selected, the jump does not occur.

\medskip
After an equilibration run, the currents are calculated as follows. We define the current of type $A$ particles between sites $i$ and $i+1$ as $j^A_i=(h^A_{R,i}-h^A_{L,i})/\tau$. Here $h^A_{R,i}$ denotes the number of hops of type $A$ particles from site $i$ to site $i+1$, while $h^A_{L,i}$ represents the number of hops of type $A$ particles from site $i+1$ to site $i$. An equivalent procedure applies to type $B$ particles. The quantity $\tau$ denotes the number of Monte Carlo steps. Finally, with two species $A$ and $B$, the total and diffusive (or relative) currents between sites $i$ and $i+1$ are $j_i=j^A_i+j^B_i$ and $j_{D,i}=j^A_i-j^B_i$, respectively. At the steady state, both currents should be the same for all consecutive $\mathcal{L}$ pairs of sites. Therefore, we can compute the average values $j=\sum_{i=0}^{\mathcal{L}-1}j_i$ and $j_D=\sum_{i=0}^{\mathcal{L}-1}j_{D,i}$.

All simulations were performed on a lattice of size $\mathcal{L}=100$, with a total number of particles $\mathcal{N} \approx 2\times 10^4$.

\subsection{Soft core interaction}

As mentioned earlier in Sec.~\ref{s.surf}, soft-core interaction admits a maximum particle number per lattice site $N_m$. The excess chemical potential is $\mu^\text{e} = -T \ln (1-\theta)$ and the thermodynamic factor is $\Gamma = 1/(1-\theta)$, with $\theta = \rho/\rho_m$ \cite{dimuro}. Using Eqs.~(\ref{e.gsc}) and (\ref{e.jthetaD}) we can show that the diffusion current, 
\begin{equation}\label{eq.jdsf}
    j_D = D_\text{id} \theta_D \partial_x \theta,
\end{equation}
is constant, since constant values of $\theta_D$ and $\partial_x \theta$ are imposed in the simulations (see Fig.\ \ref{fig:Histo}).

On the other hand, to perform the Monte Carlo simulations we need the transition rate between neighboring sites; from Eq.\ \eqref{e.WNN} we have,
\begin{equation}
W_{N_\text{R},N_\text{L}}=\nu (1-\theta_\text{L}).    
\end{equation}

\begin{figure}
    \centering
    \includegraphics[width=1\linewidth]{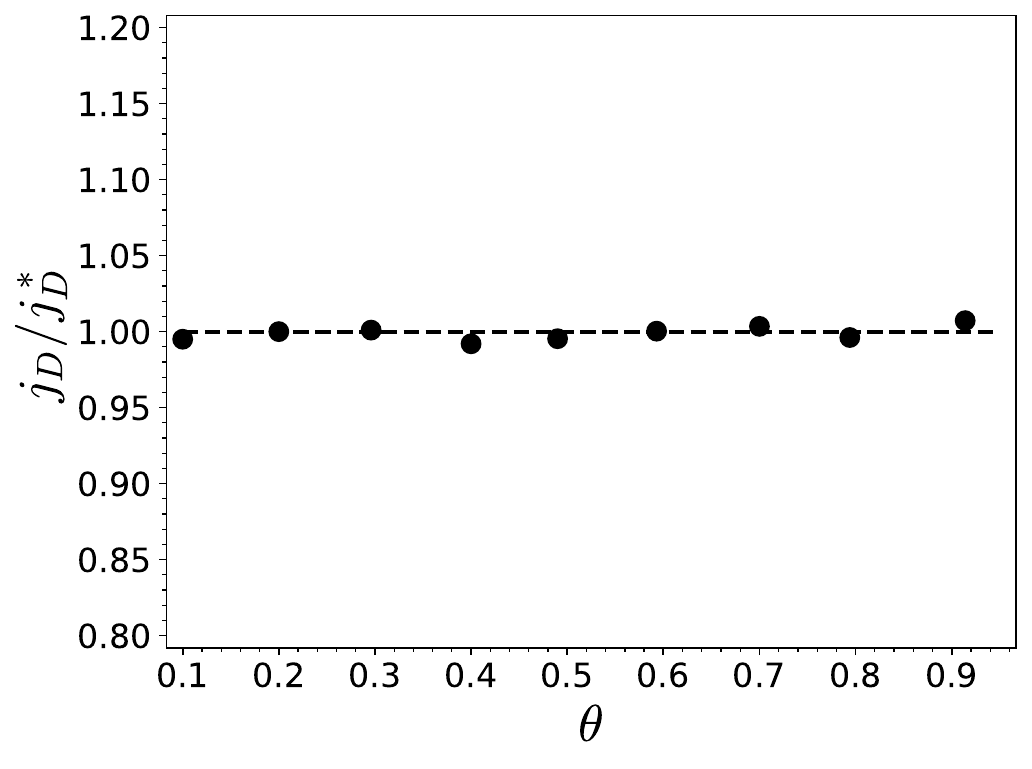}
    \caption{Diffusion current $j_D$ for soft-core interaction as a function of the coverage $\theta$, where $j^\ast_D=D_\text{id} \theta_D \partial_x{\theta}$. The dashed line represents Eq.~(\ref{eq.jdsf}) and the symbols correspond to the Monte Carlo simulation results. The statistical error of the data is less than $1\%$.}
    \label{fig:k_SF}
\end{figure}

In Fig.~\ref{fig:k_SF} we contrast Eq.~(\ref{eq.jdsf}) with the simulation results for different values of fractional coverage $\theta$, where we observe that there is a good agreement.

\subsection{Linear, quadratic and cubic excess chemical potential}

In order to verify the generality of the results, we analyze the case of an interaction such that the excess chemical potential behaves as
\begin{equation}\label{e.muex}
\beta \mu^\text{e} = \theta^k.    
\end{equation}

The thermodynamic factor $\Gamma$ and the transition rate $W$ can be obtained \cite{dimuro}:
\begin{equation}\label{e.Gam_k}
 \Gamma = 1+k \theta^k   
\end{equation}
\begin{equation}
    W_{N_\text{R},N_\text{L}}=\nu\frac{e^{(\theta^k_\text{R}-\theta^k_\text{L})/2}}{(1+k\theta^k_\text{R})^{1/2}(1+k\theta^k_\text{L})^{1/2}}.
\end{equation}

Using Eqs.~(\ref{e.jthetaD}) and (\ref{e.Gam_k}) we can calculate the diffusion current,
\begin{equation}\label{eq.jdk}
    j_D = D_\text{id} \theta_D \partial_x \theta \frac{k \theta^{k-1}}{1+k \theta ^k},
\end{equation}
which, unlike soft-core interaction, is a function of the coverage.

\begin{figure}
    \centering
    \includegraphics[width=1\linewidth]{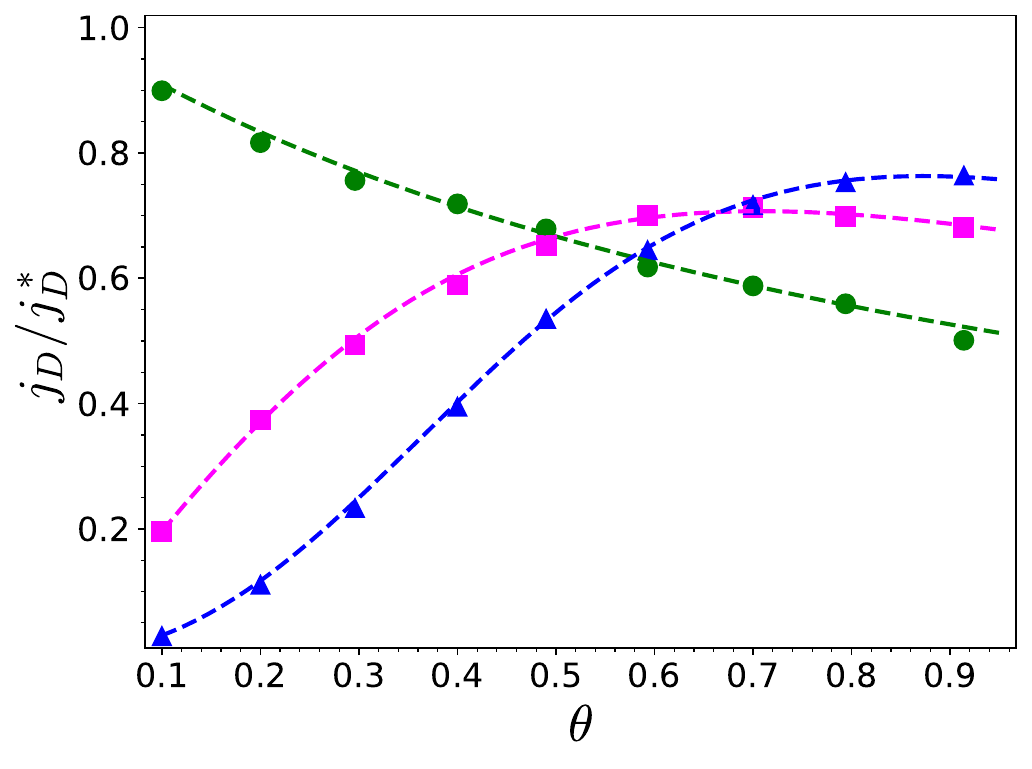}
    \caption{Diffusion current $j_D$ as a function of the coverage $\theta$, where $j^\ast_D=D_\text{id} \theta_D \partial_x{\theta}$. Linear, cuadratic and cubic chemical potentials are shown: $\mu^\text{e}=\theta^k$; $k=1$ (circles), $k=2$ (boxes), $k=3$ (triangles). The dashed line represents Eq.~(\ref{eq.jdk}) and the symbols correspond to the MC simulations. The statistical error of the data is less than $1\%$.}
    \label{fig:k_chem}
\end{figure}

In Fig.~\ref{fig:k_chem} we can see that there is a good agreement between the MC simulation results and the theoretical prediction for all depicted $k$ values.

\section{Summary and conclusions}
\label{s.conclusions}

In this work, we have developed a general thermodynamic framework for determining transport coefficients on a lattice. The main result of this study is the derivation of a general equation for the Onsager matrix ($L$) for a specified open system ensemble, where conserved quantities are exchanged with a reservoir. By establishing a simple relationship between $L$ and its ideal value ($L_\text{id}$) through the ratio of the determinants of the covariance (or Hessian) matrices, we provide a method to calculate transport properties; see Eq\ \eqref{LC}.

The derivation of the main result follows a systematic statistical thermodynamic procedure. The process begins by considering two adjacent lattice cells (L and R) in contact with a reservoir that fixes intensive parameters. The exchange of a small quantity $\vec{\delta}$ between these cells is governed by the detailed balance condition, Eq.\ \eqref{detbal}. The probabilities of the initial and final states, $P_A$ and $P_B$, are expressed using the entropy, which include non-extensive terms necessary for correctly calculating transition probabilities. By expanding the entropy difference between states up to order $O(N^{-1})$, the probability ratio $P_B/P_A$ is determined, see Eq.\ \eqref{pratio}. To eliminate unknown functions in the transition rate equation, an ideal reference system is introduced where particle interactions are neglected. The transition rate of the real system is then expressed as the ideal transition rate scaled by an excess factor, $e^{-2\Delta S_e}$, where $\Delta S_e$ represents the excess part of the non-extensive entropy. A key step in the derivation is expressing this excess entropy term in terms of the covariance matrix. Specifically, it is known that $2\Delta S_e = -\ln(\det(C)/\det(C_\text{id}))$ \cite{hoyuelos}. Finally, by considering a small departure from equilibrium driven by thermodynamic forces (gradients of intensive variables), the net current is calculated. Comparing this to the phenomenological linear relation $\vec{J} = -L \cdot \vec{F}$ yields the central result: $L = \frac{\det(C)}{\det(C_\text{id})} L_\text{id}$.

This procedure demonstrates that the transport coefficients of an interacting system can be determined solely from its thermodynamic equation of state and the transport properties of a corresponding ideal system.

This general result was successfully used to reproduce the transport coefficients in two classic diffusion frameworks. Specifically, the derivation was applied to the Darken model for substitutional diffusion in solids, correctly identifying the role of the thermodynamic factor in the relationship between intrinsic and tracer diffusion. Furthermore, the approach was used to obtain the non-diagonal diffusion matrix of the Zhdanov model for surface diffusion of Langmuir particles. These applications demonstrate that the thermodynamic encoding of interactions within the chemical potential, combined with the knowledge of the transport matrix in the ideal case, provide a versatile description of diffusion processes with interactions.

In summary, the main novel insights of this work lie in establishing a general theoretical framework for calculating the  Onsager matrix $L$, a formulation that until now had been verified for the Darken equation and the Zhdanov model. Furthermore, our approach connects transport theory with thermodynamic geometry: the presence of the Hessian matrix determinant in the relation for $L$ provides a geometric perspective, as this determinant serves as a measure of information density on the state-space manifold, reflecting how interparticle interactions distort the underlying thermodynamic landscape to modify macroscopic transport.

Finally, it is worth addressing some challenges associated with determining the ideal diffusivity, $D_{\text{id}}$. Our framework assumes $D_{\text{id}}$ to be known and determines how interactions modify transport. However, in real systems, such as collective surface diffusion on crystals, the energy landscape can be highly complex, featuring multiple adsorption states and varying energy barrier heights. In such cases, even in the absence of direct particle-particle forces, the presence of neighboring particles may induce indirect interactions by altering the potential energy surface itself, rendering $D_{\text{id}}$ concentration-dependent. Obtaining $D_{\text{id}}$ under these complex conditions remains a challenging problem (see, for example, \cite{minkowski}); nevertheless, once $D_{\text{id}}$ and the equation of state are established, the general thermodynamic approach developed here determines the transport properties.

\begin{acknowledgments}
	This work was partially supported by Universidad Nacional de Mar del Plata (UNMdP, Argentina, Project No.\ 15/E1155).
\end{acknowledgments}

\bibliography{tpdif.bib}

\end{document}